\documentclass[12pt,titlepage]{article}

\usepackage{natbib}
\usepackage{multicol}

\usepackage{lscape}
 
\usepackage{subfig}
\usepackage{color}
\usepackage{amstext}
\usepackage{amsmath, amsthm, amssymb}
\usepackage{amsfonts}
\usepackage{graphicx}

\begin{document}

\baselineskip=18pt


\title{Regional Probabilistic Fertility Forecasting by Modeling Between-Country Correlations}
\author{Bailey K. Fosdick and Adrian E. Raftery
\thanks{Bailey K. Fosdick is a Graduate Research Assistant and Adrian E. Raftery is a Professor of Statistics and Sociology, both at the Department of Statistics, Box 354322, University of Washington, Seattle, WA 98195-4322 (Email: bfosdick@u.washington.edu/raftery@u.washington.edu). This work was supported by NIH grants R01 HD054511 and R01 HD070936. 
The authors thank Leontine Alkema, Sam Clark and Patrick Gerland 
for helpful comments and discussion.}  \\
Department of Statistics \\
University of Washington }
\date{\today}
\maketitle

\begin{abstract}
The United Nations (UN) Population Division 
is considering producing probabilistic projections for the total fertility rate (TFR) using the Bayesian hierarchical model of \cite{Leontine}, which produces predictive distributions of TFR for individual 
countries.  The UN is interested in publishing probabilistic projections 
for aggregates of countries, such as regions and trading blocs. 
This requires joint probabilistic projections of future country-specific
TFRs, taking account of the correlations between them.  We propose an extension of the Bayesian hierarchical model that allows for probabilistic projection of TFR for any set of countries.  We model the correlation between country forecast errors as a linear function of time invariant covariates, namely whether the countries are contiguous, whether they had a common colonizer after 1945, and whether they are in the same UN region.  The resulting correlation model is incorporated into the Bayesian hierarchical model's error distribution.  We produce predictive distributions of TFR for 1990-2010 for each of the UN's primary regions.  We find that the proportions of the observed values that fall within the prediction intervals from our method are closer to their nominal levels than those produced by the current model.  Our results suggest that a significant proportion of the correlation between forecast errors for TFR in different countries is due to countries' geographic proximity to one another, and that if this correlation is accounted for, 
the quality of probabilitistic projections of TFR for regions and other 
aggregates is improved.  
\end{abstract}

\section{Introduction}

The United Nations (UN) Population Division produces population estimates and projections every two years for all countries and publishes them in the biennial \textit{World Population Prospects} (WPP).  These projections are used by UN agencies and governments
for planning, monitoring development goals and as inputs to climate change
and other models. They are also widely used by social and health science 
researchers and the private sector.
The UN produces these population forecasts by projecting countries' age-
and sex-specific fertility, mortality, and migration rates, and combining them to obtain age- and sex-specific population 
sizes using the standard cohort component method.

In this paper, we focus on the fertility component.  Country fertility in a given time period is summarized by the period total fertility rate (TFR), which is the average number of children a woman would bear if she lived past the end of the reproductive age span and at each age experienced the age-specific fertility rates of the given country and time period.  
Projections of future TFR are decomposed using forecasted age schedules to obtain projections of age-specific fertility rates.
 
The WPP reports three projection variants (low, medium, and high) for the population and vital rates based on expert opinion and models of historical patterns.
The low and high variants correspond to TFR half a child below and above the medium value, respectively.  A drawback of these projections is that the range given by the low and high variants has no probabilistic interpretation and hence does not reflect the uncertainty in the forecasts.  
 
For the 2010 WPP \citep{WPP}, the UN used as its medium projection
the predictive median of TFR 
from a Bayesian hierarchical model developed by \cite{Leontine}.  
We refer to this model as the ``current model''.
This model produces predictive probability distributions of each country's 
TFR, although the distributions were not used in the 2010 WPP.
The model is based on the demographic transition, where countries move from high birth and death rates to low birth and death rates, and is composed of three phases: before, during and after the fertility transition.  Predictions from this model are typically summarized by the median country TFR prediction and the 80\% or 95\% prediction interval.

\begin{figure}
\center
\includegraphics[scale=.6]{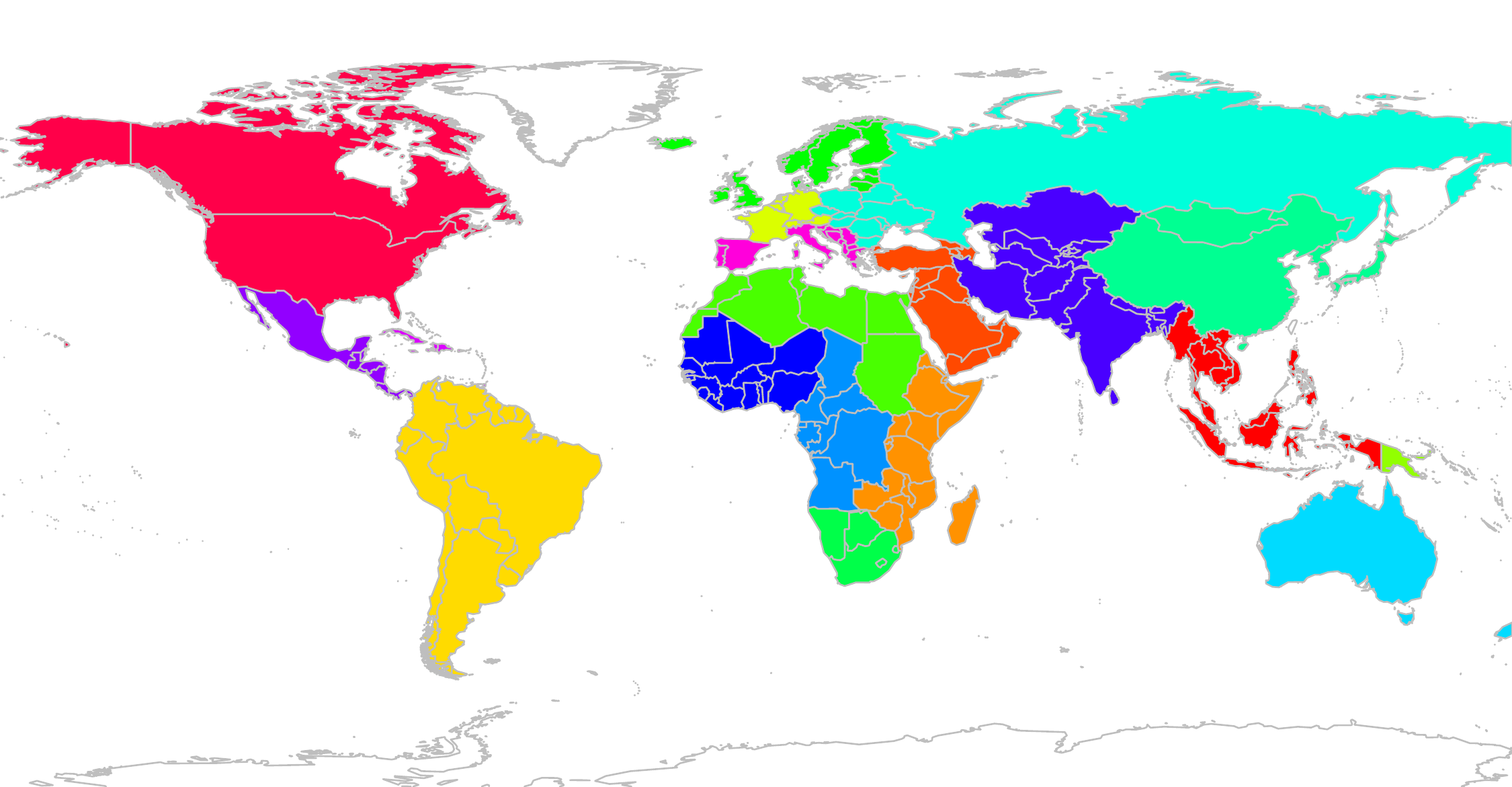}
\caption{Primary regions of the world as identified by the UN.}
\label{map}
\end{figure}

In addition to producing population estimates at the country level, the UN also provides projections for country aggregates such as geographic regions and trading blocs.  
The country TFR projections from the current Bayesian hierarchical model of \cite{Leontine} can be combined to obtain regional probabilistic TFR projections, 
provided the current model takes account of the dependence between countries' fertility rates.  However, if dependence exists between country TFRs that is not accounted for in the Bayesian hierarchical model, treating the country-specific projections as independent may underestimate the uncertainty about the future TFRs and populations 
of aggregates.  

Figure \ref{map} shows the UN's 22 primary regions of the world and Table \ref{indonly} summarizes the coverage probability of the out-of-sample TFR prediction intervals for these regions based on the current model.  
The coverage probabilities of the region-specific predictive intervals
are smaller than the nominal levels, even though the country-specific 
coverages have been found to be approximately
correct \citep{Leontine}.  This suggests that the assumption of 
independent forecast errors may not be appropriate.

\begin{table}[ht]
\begin{center}
\small
\caption{Proportion of observed regional TFRs that fall within the specified out-of-sample prediction intervals obtained from the current Bayesian hierarchical model of \protect\cite{Leontine}.}
\label{indonly}
\begin{tabular}{c|rrr}
Time Period & 80\% CI & 90\% CI & 95\% CI  \\ 
  \hline \hline
1990-1995 & 0.73  & 0.86  &  0.95  \\ 
1995-2000 & 0.68  & 0.73  &  0.86  \\ 
 2000-2005 & 0.59 &  0.73  & 0.82    \\ 
 2005-2010  &  0.73&  0.82 &  0.91  \\ 
    \hline
All &  0.68&  0.78&  0.89  \\ \hline
\end{tabular}
\end{center}
\end{table}

In this article, we propose an extension to the Bayesian hierarchical model that produces TFR estimates for any aggregate of countries by modeling the residual correlation between country TFRs.  Our extension adds a correlation structure to the error distribution in the hierarchical model, where the correlation between a pair of countries is modeled as a linear function of time invariant covariates.  Three covariates are chosen: whether two countries are contiguous, whether they had a common colonizer after 1945, and whether they are in the same UN region.  This model provides estimates of the correlation between any pair of countries, even when empirical estimates are not available.  

Correlation matrices based on the linear function of covariates in our extension are not positive semidefinite, and hence are not valid for prediction, for many values of the covariate coefficients.  This makes estimation of the covariate coefficients difficult.  In addition, while simulation of forecast errors from a correlation matrix requires only that the matrix be positive semidefinite (and so may be singular), traditional estimation procedures such as maximum likelihood estimation do not accommodate singular covariance matrices.  Thus, such estimation methods are unsatisfactory for this problem.  We propose instead estimating the coefficients by maximizing a pseudo-likelihood function defined by the country pairwise correlations.

This paper is organized as follows.  In the next section we review the current model and introduce the correlation model extension.  
We also describe the exploratory analyses that led to the choice of model
extension.
An estimation procedure based on a pseudo-likelihood function is described, and model validation results are then presented for the prediction of the TFR in each of the UN's regions.  We show theoretically which regional prediction intervals are most affected by the correlation model and compare the pairwise country correlation values from our model and those obtained in previous studies.  We conclude with a discussion of previous work. 

\section{Methodology}

\subsection{Current Model}

The Bayesian hierarchical model of \cite{Leontine} divides the evolution of
TFR in a country into three phases: before, during and after the
fertility transition.
During the fertility transition, the TFR for country $c$ in time period $t$, $f_{c,t}$, is modeled as following a systematic decline curve with normally distributed random errors.  After the fertility transition is complete, the TFR is modeled as a first order autoregressive process that ultimately fluctuates about 2.1, which is considered replacement level fertility.  If $\boldsymbol{f_{t}} = (f_{1,t},....,f_{C,t})$ is the TFR for all countries at time $t$, the model can be written as follows:

 \begin{equation} 
\boldsymbol{f_{t}} = \boldsymbol{m_{t}}  + \boldsymbol{\epsilon_{t}}, \hspace{.4in} \boldsymbol{\epsilon_{t}} \sim \text{N}(0, \Sigma_{t}=\boldsymbol{\widetilde{\sigma}_{t}^{T}} \boldsymbol{\widetilde{\sigma}_{t}})   \label{curmodel} 
\end{equation} 

\noindent \begin{multicols}{2}
\indent Fertility transition phase:\\
\indent$ m_{c,t}  =  f_{c,t-1} - d(\boldsymbol{\theta_{c}},f_{c,t-1})$ \\
\indent$\widetilde{\sigma}_{c,t}  =  \sigma_{c,t}(\boldsymbol{\theta_{c}},f_{c,t-1})$ \\

\noindent Post-transition phase:  \\
$m_{c,t}  = 2.1 + 0.9(f_{c,t-1} - 2.1)  $\\
$ \widetilde{\sigma}_{c,t}  = s = 0.2  $
 \end{multicols}

In (\ref{curmodel}), the quantities in bold font are vectors
whose elements correspond to different countries, 
$d(\boldsymbol{\theta_{c}},f_{c,t-1})$ is a double logistic function controlling the rate of the fertility decline, and $\boldsymbol{\theta_{c}}$ is a vector of country-specific parameters.  The expected TFR in the next time period, $m_{c,t}$, and the variances of the random errors, $\widetilde{\sigma}_{c,t}^{2}$, differ in the transition and post-transition phases.  Since a country's TFR is not modeled before the fertility decline, the vector $\boldsymbol{f_{t}}$ for any time point $t$ contains only those countries that have started or completed their fertility transition.   
 
\begin{figure}
\begin{center}
\hspace*{-.2in} \includegraphics[scale=.8]{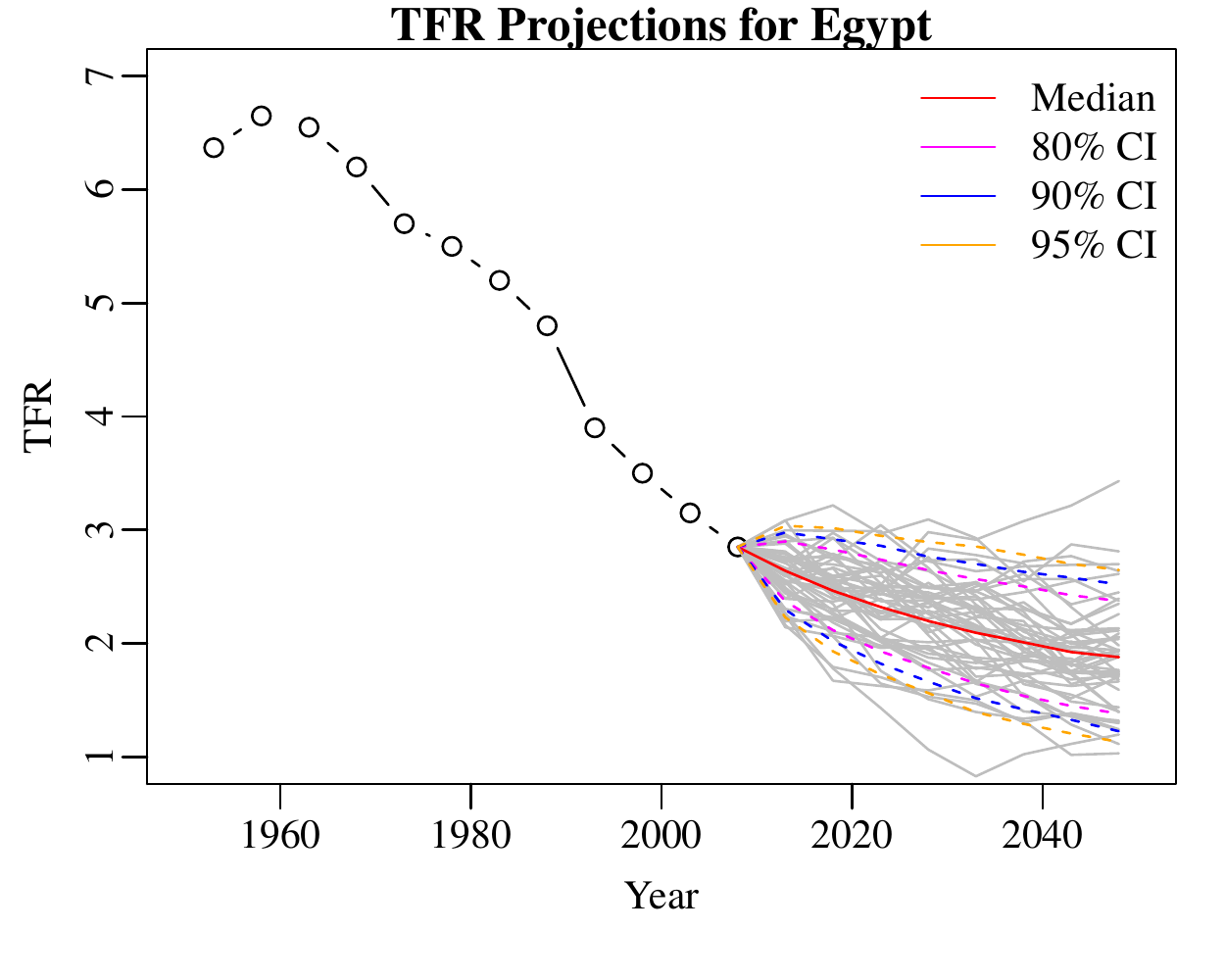}
\end{center}
\caption{The plot shows TFR projections (grey) and probabilistic prediction intervals for Egypt from 2010 to 2050 from the Bayesian hierarchical model. }
\label{expTFR} 
\end{figure}

The data used to estimate the country parameters $\boldsymbol{\theta_{c}}$ in the current model are the five-year time period TFR estimates from 1950 to 2010 in the 2010 WPP.  A posterior distribution of the parameters is produced which indicates the probable values of the parameters given the data.  In addition, a predictive distribution of TFR values for each country can be obtained by forecasting future values using the relations in \eqref{curmodel}.  
Figure \ref{expTFR}    
shows the predictive distribution of TFR for Egypt from 2010 to 2050.  This distribution is summarized by the median prediction and the 80\%, 90\%, and 95\% prediction intervals.  

\subsection{Correlation Model}

As discussed above, the regional TFR prediction intervals from the current model tend to be too narrow (see Table \ref{indonly}).  This suggests there is excess correlation between countries' TFRs that is not accounted for in the current model.  To capture this excess correlation, we propose modifying the error structure in \eqref{curmodel} to allow for correlation between countries as follows:
\begin{align}
&\boldsymbol{\epsilon_{t}} \sim \text{N}(0, \Sigma_{t}=\boldsymbol{\widetilde{\sigma}_{t}^{T}}R_{t} \boldsymbol{\widetilde{\sigma}_{t}}).  \label{corerr} 
\end{align}
The $(i,j)$ element of the matrix $R_{t}$ is the correlation between the TFR forecast errors (i.e. deviations from the mean predicted values $m_{c,t}$) for country $i$ and country $j$ in time period $t$.  

Our exploratory analyses, described below, indicated that the correlations 
had a different pattern when both countries had low fertility than 
otherwise, and our model allows for this. We sought to model
the correlations using temporally stable characteristics of the country pairs,
so that they could reasonably be used for projections.
Thus, the elements of the correlation matrix are modeled as follows:
\begin{align}
& R_{t}[i,j] = \left\{
     \begin{array}{ll}
      1 &\hspace{.1in} \mbox{if }  i = j , \\
       \rho^{(1)}_{ij}  & \hspace{.1in} \mbox{if }  f_{i,t-1}<\kappa \text{  and  } f_{j,t-1}<\kappa , \\
       \rho^{(2)}_{ij}  & \hspace{.1in}  \mbox{if } f_{i,t-1}\ge\kappa \text{  or  } f_{j,t-1}\ge \kappa , \\
     \end{array}
   \right.  \label{piecewisecor}   \\[4pt]
& \rho_{i,j}^{(k)} =  \beta^{(k)}_{0} + \beta^{(k)}_{1}\text{contig}_{i,j} + \beta^{(k)}_{2}\text{comcol}_{i,j} + \beta^{(k)}_{3}\text{sameRegion}_{i,j} \hspace{.2in} \text{for }k \in \{1,2\}, \; i \not=j , \notag
\end{align}
where contig$_{i,j}=1$ if countries $i$ and $j$ are contiguous and 0 if not,
comcol$_{i,j}=1$ if they had a common colonizer after 1945, 
and sameRegion$_{i,j}=1$ if they are in the same UN region.  

The correlation model in (\ref{piecewisecor}) states that when countries $i$ and $j$ both have TFR below $\kappa$, the correlation of their errors in the next time period is $\rho^{(1)}_{ij}$, and the correlation is $\rho_{ij}^{(2)}$ when at least one of them has a TFR greater than $\kappa$.  
In both cases, the correlation between two countries is modeled as a 
linear combination of the three pairwise country covariates.  
The parameters to be estimated therefore include the threshold $\kappa$, 
$\{\beta^{(1)}_{0}, \beta^{(1)}_{1}, \beta^{(1)}_{2}, \beta^{(1)}_{3}\}$ for the correlation when both countries have TFR less than $\kappa$, and $\{\beta^{(2)}_{0}, \beta^{(2)}_{1}, \beta^{(2)}_{2}, \beta^{(2)}_{3}\}$ for the correlation when at least one of the two TFRs is greater than $\kappa$.

Since the diagonal elements  of $R_{t}$ are equal to one, the joint predictive distribution of all country TFRs will have the same country marginal predictive distributions as those from the current model. Thus expanding the model
to allow for correlation will not change the marginal country-specific
predictive distributions, which is desirable given the good performance
of the current model for individual countries.

\subsection{Exploratory Analysis}

Exploratory analysis of one-time-period-ahead forecast errors from the model 
of \cite{Leontine}
and WPP data from 1950 to 2010 guided specification of the correlation model structure.  For each time period and country, the forecast error is the difference between the observed TFR and the average predicted value given TFR in the previous time period.  
Estimating the correlations between these forecast errors is difficult because the estimates are based 
on a small amount of data (at most 11 five-year periods), and because the
country-specific predictive means and variances are given by the 
Bayesian hierarchical model.
To obtain empirical estimates of the correlation between the forecast errors for two countries,
conditional on their predictive variances, we used the posterior mean with an arc-sine prior.  This estimator was proposed by \cite{CorPaper}, 
who showed it to have good small sample performance compared to other frequentist and Bayesian estimators.  

\begin{figure}
\begin{center}
\includegraphics[scale=.7]{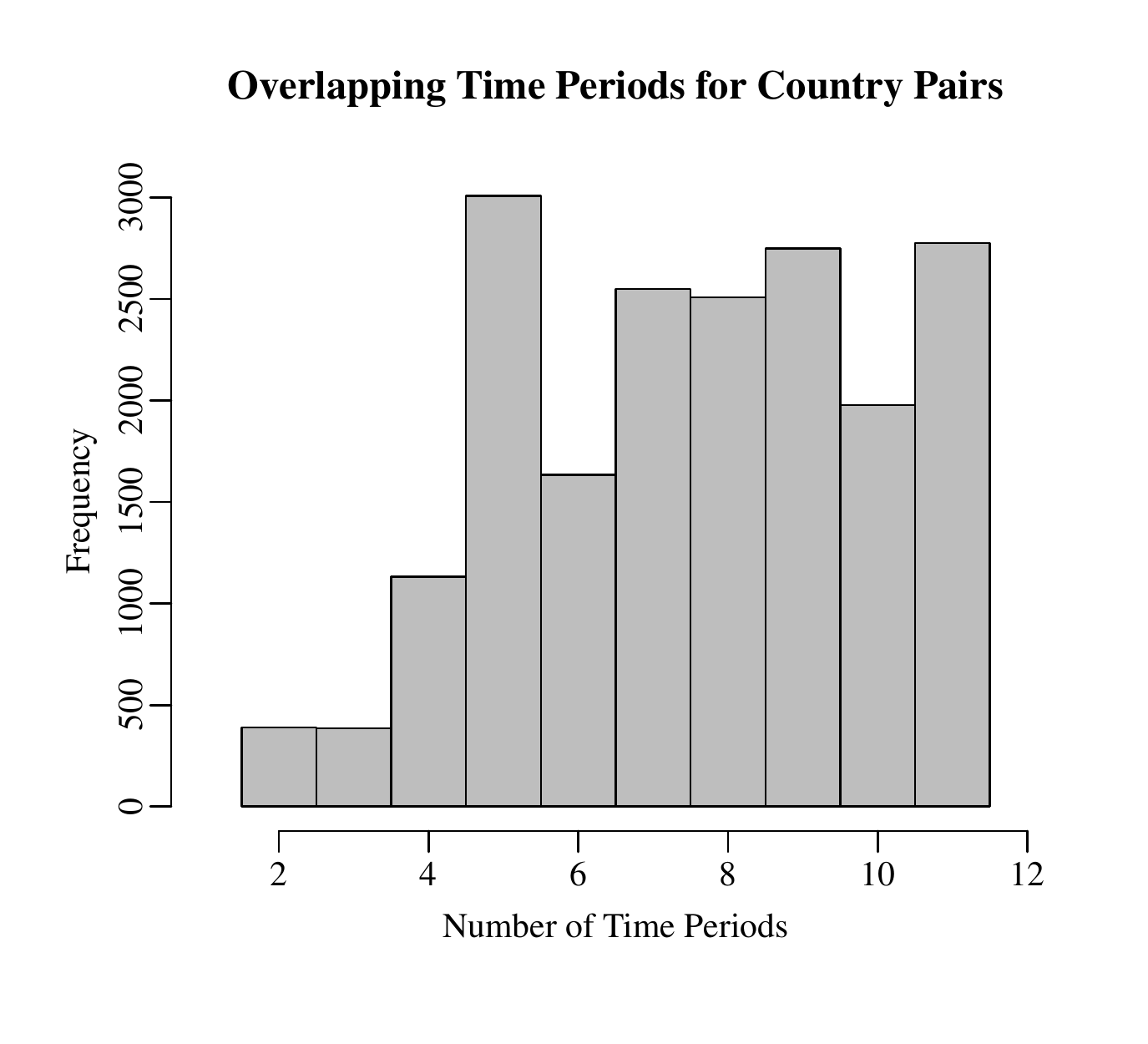} 
\end{center}
\caption{Number of 5-year time periods from 1955 to 2010 for each country pair after the start of both country's fertility declines.}
\label{numobs} 
\end{figure}

Figure \ref{numobs} shows the number of five-year time periods from 1955 to 2010 for each country pair after both had started their fertility decline.  These counts represents the number of forecast errors used to compute each correlation estimate.  Since a number of countries have only recently started their fertility decline, many pairwise correlation estimates were based on only a few observations or, in the case of only two overlapping time periods, were not computed at all.  We therefore modeled the correlation structure rather than directly 
using the noisy empirical estimates from the raw data.

The correlation estimates were higher on average when both countries
had low TFR and had completed or nearly completed the fertility transition.
This led us to specify one model for the correlation when
the TFRs of both countries were below a threshold $\kappa$, 
and a different model when at least one country had TFR above $\kappa$, 
where $\kappa$ is to be estimated from the data. 

The average estimated correlation between countries in the same UN-defined 
region when both have TFR below 3 was 0.37, and for countries in different 
regions was 0.09, using only correlation estimates based on at least 
eight time periods.  
This suggests that the correlation between forecast errors at low TFR levels
may be related to geographical proximity, and motivates modeling the 
correlation as a function of geographical predictors.

Since our aim is to make long-term projections,
we consider only predictors that are essentially time invariant.
A database of country pairwise covariates is available from the 
Centre d'Etudes Prospectives et d'Informations Internationales (CEPII) 
\citep{CEPII}.  A list of the pairwise covariates in this database is shown in Table \ref{vars}.   
Using linear regression,
we found that the covariates most predictive of the correlation estimates are whether two countries are contiguous (contig), whether they share a common colonizer after 1945 (comcol), and whether they are in the same UN region (sameRegion).  

\begin{table}[ht]
\begin{center}
\caption{Pairwise variables in the CEPII database \citep{CEPII}.}
\label{vars}
\begin{tabular}{ll} \hline 
contiguous (contig) &common official language \\
common colonizer after 1945 (comcol) \hspace{.4in}& share a language spoken by at least 9\%  \\ 
colonial link &  geodesic distance by most important cities  \\
colonial relationship after 1945 &geodesic distance by capital cities  \\ 
currently in a colonial relationship & distance weighted by city populations: arithmetic mean \\
were/are the same country &distance weighted by city populations: harmonic mean
\\  \hline
      \end{tabular}
\end{center}
\end{table}

\section{Parameter Estimation}\label{sec:est}

Our method for estimating the parameters of the correlation model in
(\ref{piecewisecor}) relies on
the one-time-period-ahead standardized forecast errors.  The Bayesian hierarchical model of \cite{Leontine} was fit to the 2010 WPP TFR estimates from 1950 to 2010, and posterior distributions of $\boldsymbol{\theta_{c}}$ given the data were obtained for each country.  Using these parameter estimates and the TFR in a given time period, a predictive distribution of the expected TFR $m_{c,t}$ for the next time period was computed.  
The value of $m_{c,t}$ for a parameter vector $\boldsymbol{\theta_{c}}$ is
 \begin{displaymath}
   \hat{m}_{c,t}| f_{c,t-1}^{WPP},\boldsymbol{\theta_{c}}   = \left\{
     \begin{array}{ll}
       f_{c,t-1}^{WPP} - d(\boldsymbol{\theta_{c}},f_{c,t-1}^{WPP}), &  \text{during the fertility transition,}\\
        2.1 + \rho(f_{c,t-1}^{WPP}  - 2.1), &  \text{after the fertility transition,}
            \end{array}
   \right.
\end{displaymath} 
where $f_{c,t-1}^{WPP}$ is the 2010 WPP TFR estimate for country $c$ at time period $t-1$.  For each sample $\boldsymbol{\theta^{(k)}_{c}}$ from the posterior distribution, there is a corresponding expected TFR value at time $t$.  

We define the parameter-specific standardized forecast error $e_{c,t}(\boldsymbol{\theta_{c}})$ for country $c$, time period $t$, and parameter vector $\boldsymbol{\theta_{c}}$ as
\begin{displaymath}
e_{c,t}(\boldsymbol{\theta_{c}}) = \frac{f_{c,t} - E[f_{c,t}|f_{c,t-1},\boldsymbol{\theta_{c}}]}{SD[f_{c,t}|f_{c,t-1},\boldsymbol{\theta_{c}}]} = \frac{f_{c,t}^{WPP} - \hat{m}_{c,t}| f_{c,t-1}^{WPP},\boldsymbol{\theta_{c}}}{\widetilde{\sigma}_{c,t}}.\label{stderr}
\end{displaymath}
We define the standardized one-time-period-ahead forecast error $\overline{e}_{c,t}$ as
the average over the posterior samples of the parameter-specific 
forecast errors, namely
$$ \overline{e}_{c,t} = \frac{1}{K} \sum_{k=1}^{K} e_{c,t}(\boldsymbol{\theta_{c}}^{(k)}) , $$
where $K$ is the number of parameter samples from the posterior distribution.  

The standardized errors can be viewed as samples from a multivariate 
normal model with correlation matrix $R_{t}$, 
$\boldsymbol{\overline{e}_{t}}( \boldsymbol{\theta})  \sim \text{N} \left(\boldsymbol{0}, R_{t} \right)$ for  $t=1955,...,2010$.  
Ideally we would estimate the correlation model parameters $\{\kappa$, $\beta^{(1)}_{0}$, $\beta^{(1)}_{1}$, $\beta^{(1)}_{2}$, $ \beta^{(1)}_{3}$, $\beta^{(2)}_{0}$, $ \beta^{(2)}_{1}$, $ \beta^{(2)}_{2}$, $ \beta^{(2)}_{3}\}$ via maximum likelihood estimation based on the multivariate normal model.
However, this is made challenging by the fact that 
for any time period $t$, the vector $\boldsymbol{\overline{e}_{t}}$ contains standardized errors for only those countries that have started their fertility decline by time $t$, and that for many parameter values the estimated correlation matrix is not positive definite, making the likelihood undefined.

Instead, we took a pseudo-likelihood approach that approximates the 
multivariate normal likelihood by a product of bivariate normal likelihoods 
\citep{Besag1975}.  We call this the \textit{Aggregation Pseudo-Likelihood (APL)} and define it as
\begin{align}
\text{L}_{\text{APL}}( \kappa, \boldsymbol{\rho^{(1)}},\boldsymbol{\rho^{(2)}}|\boldsymbol{\overline{e}}) = \prod_{t=1}^{T} \prod_{i < j}& \bigg[ \text{L}_{1}(\rho^{(1)}_{ij}|\overline{e}_{i,t},\overline{e}_{j,t}) \cdot \text{I}\left[(f_{i,t-1}<\kappa) \cap (f_{j,t-1}<\kappa)\right] \notag \\
 &+   \text{L}_{2}( \rho^{(2)}_{ij}|\overline{e}_{i,t},\overline{e}_{j,t}) \cdot  \text{I} \left[(f_{i,t-1}\ge\kappa) \cup (f_{j,t-1}\ge\kappa) \right] \bigg] ,
\label{pseudoLL}
 \end{align}
where $T$ is the number of observed time periods and L$_{1}$ and L$_{2}$ are bivariate normal likelihoods with zero means, variances equal to one, and correlations $\rho^{(1)}_{ij}$ and $\rho^{(2)}_{ij}$, respectively.

Using the APL, the likelihood can be maximized separately over 
$\{\beta^{(1)}_{0}, \beta^{(1)}_{1}, \beta^{(1)}_{2}, \beta^{(1)}_{3}\}$ 
and $\{\beta^{(2)}_{0}, \beta^{(2)}_{1}, \beta^{(2)}_{2}, \beta^{(2)}_{3}\}$ 
for a fixed value of $\kappa$.
For each value of the threshold $\kappa$ from 0.5 to 9 children
at intervals of 0.1, we estimated the model parameters
by maximizing the APL in (\ref{pseudoLL}) numerically using a Nelder-Mead method.

The APL was maximized at $\kappa=5$ children, and the corresponding 
regression coefficients are shown in Table \ref{params}.  
These estimates mirror the exploratory analysis result that correlations
are larger on average when both countries have lower TFR.  
At TFR values below $\kappa$, 
the correlation between two countries that are contiguous and in the same region but do not share a common colonizer is 0.46.  
The correlation between countries that are not in the same region 
but are contiguous is 0.37.  The corresponding values when at least one TFR 
is above $\kappa$ are 0.13 and 0.11, respectively.  
Country pairs with no colonial or geographic relationship have a correlation of 
0.11 when both TFRs are below $\kappa$, and 0.05 otherwise.  
This illustrates that the correlation between two countries is associated
with their geographic and colonial relationship.

 \begin{table}[ht]
\begin{center}
\caption{Parameter estimates for the correlation model
(\ref{piecewisecor}).  The estimate of the threshold $\kappa$ is 5.}
\begin{tabular}{l|cccc}
& intercept & contig & comcol & sameRegion \\
& ($\beta_{0}$) & ($\beta_{1}$) & ($\beta_{2}$) &  ($\beta_{3}$) \\
 \hline
Both country TFRs below $\kappa$ & 0.11 & 0.26 & 0.05 & 0.09 \\
 At least one country TFR greater than $\kappa$ & 0.05 &  0.06 & 0.00 & 0.02
\\ \hline
\end{tabular}
\label{params}
\end{center}
\end{table}

For many time periods, the APL estimates of the parameters result in 
estimated correlation matrices $R_{t}$ that are symmetric but not positive 
semidefinite.  However, the correlation matrix must be positive semidefinite to use it for simulation of forecast errors.
The symmetric positive semidefinite matrix closest in Frobenius norm to a given symmetric matrix is obtained by zeroing out all negative eigenvalues of the original matrix \citep{NearestCov} and then reconstructing
the matrix.
Thus, at each time point $t$ for which $R_{t}$ is not positive semidefinite, we perform the following procedure:
\begin{enumerate}
\item
Compute the eigenvalue decomposition of $R_{t}$ to express it as $R_{t}=UDU^{T}$, where $U$ is an orthogonal matrix of eigenvectors and $D$ is a diagonal matrix of eigenvalues.  
\item
Replace all the negative eigenvalues by zero. The matrix $D$ is thereby
changed to $\widetilde{D}$.
\item Compute the reconstructed matrix 
$\widetilde{R}_{t} = U\widetilde{D}U^{T}$.
\item
The diagonal elements of $\widetilde{R}_{t}$ will not equal one unless the original matrix was positive semidefinite.  Therefore, treat $\widetilde{R}_{t}$ as a covariance matrix and rescale it to obtain a reconstructed and rescaled correlation matrix $\widehat{R}_{t}$ to use in the projections.
\end{enumerate}

The rescaling of the correlation matrix in step 4 ensures that the 
predictive distribution of TFR for any individual country 
remains the same as from the current Bayesian hierarchical model of 
\cite{Leontine}.  Thus, only joint predictive distributions of the TFRs
in more than one country are affected.
Note that the matrix approximation $\widehat{R}_{t}$ that results 
from this procedure is singular unless the original matrix $R_{t}$ 
is positive definite, but the predictive distributions remain well defined.

 \section{Results}\label{sec:res}
   
\subsection{Model Validation}

We assessed the model by estimating the current hierarchical model parameters from the data for 1950 to 1990,
projecting regional TFR for the UN's 22 primary regions from 
1990 to 2010 using the error correlation structure, and comparing the probabilistic projections
with the actual observations for the four held-out five-year periods.
We approximated regional TFR by a weighted average of country-specific
TFRs, with weights proportional to the current female populations
of each country. A similar approximation was used for regional 
life expectancy by Raftery et al (2012a). \nocite{Chunn}

Posterior distributions of the Bayesian hierarchical model parameters were obtained based on data from 1950 to 1990.  The parameter values in Table \ref{params} were used in the correlation model and not re-estimated based on the restricted data set, since these estimates are already based on very limited data.
Projections of TFR were obtained for the four five-year time periods from 1990 to 2010 under the current model assuming independent errors and using our proposed
error correlation structure.  The predictive distributions of the weighted average TFR for each of the 22 regions were compared to the observed weighted average values.  

Table \ref{prop} shows the proportion of observed weighted averages that fell within the 80\%, 90\%, and 95\% prediction intervals from both approaches.  
In each case the observed proportion was closer or as close to the 
theoretical value under the correlation model than under the independence 
model.

\begin{table}[ht]
\begin{center}
\small
\caption{Proportion of observed regional weighted average TFRs that fall within the specified prediction intervals.}
\label{prop}
\begin{tabular}{ccrrr}
  \hline
Time Period & Model & 80\% CI & 90\% CI & 95\% CI  \\ 
  \hline \hline
1990-1995& Independence & 0.73  & 0.86  &  0.95  \\ 
& Correlation &0.86 & 0.91 & 0.95 \\
&&&& \\
1995-2000& Independence & 0.68  & 0.73  &  0.86  \\ 
& Correlation &0.73 & 0.86 & 0.95 \\
&&&& \\
 2000-2005 & Independence &0.59 &  0.73  & 0.82    \\ 
& Correlation& 0.64 &  0.73&  0.95 \\
&&&& \\
 2005-2010 & Independence &  0.73&  0.82&  0.91  \\ 
& Correlation &  0.77 & 0.86 & 0.91 \\
\noalign{\medskip}
    \hline
    \noalign{\medskip}
All & Independence &  0.68&  0.78&  0.89  \\ 
& Correlation &  0.75 & 0.84 & 0.94 \\    \hline
\end{tabular}
\end{center}
\end{table}
 
\begin{figure}
\center
\includegraphics[scale=.75]{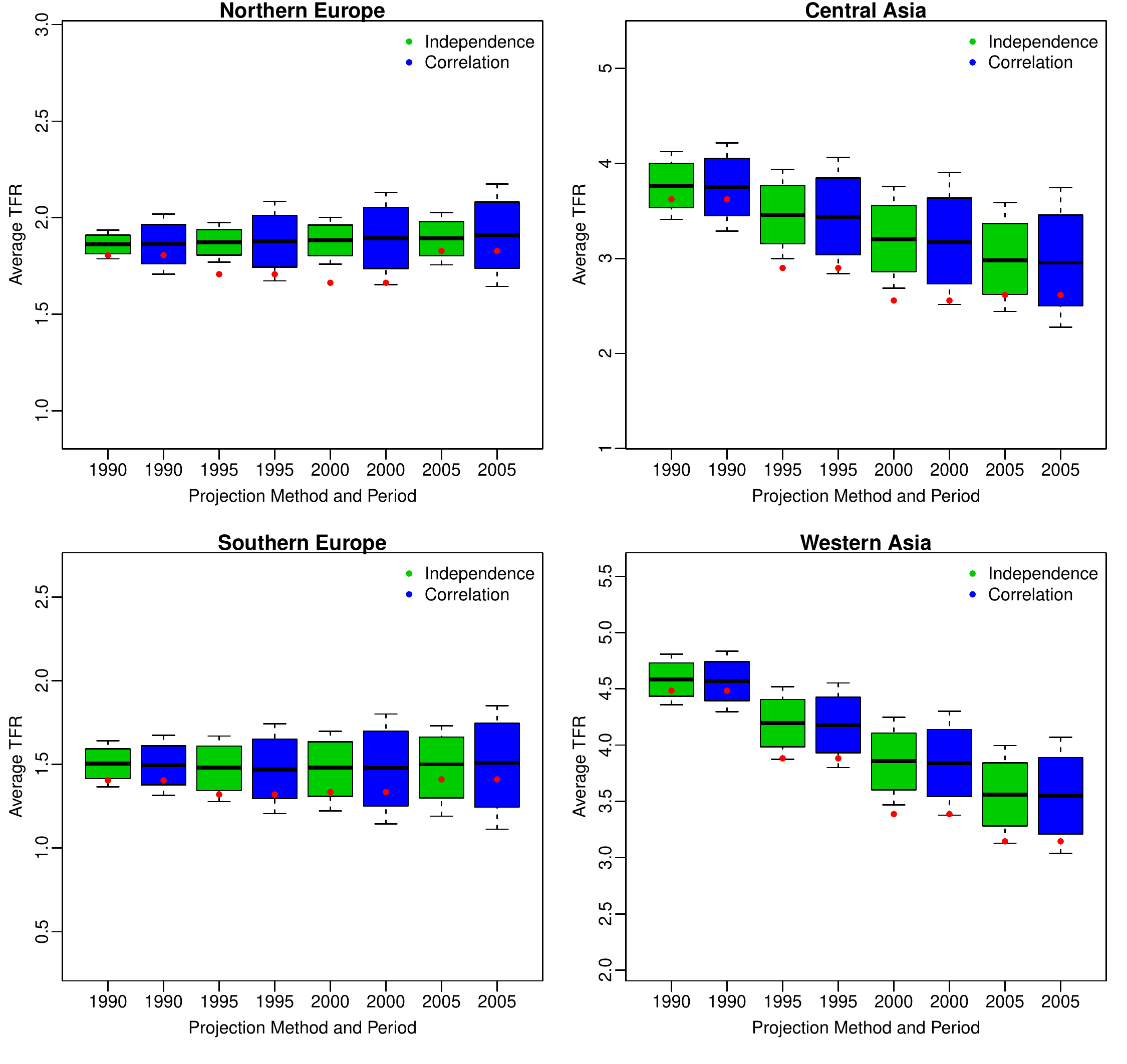}
\caption{Boxplots showing the 80\% and 95\% prediction intervals for the regional weighted average TFR for the current model assuming independent errors and that with the correlation error structure.  The box of each boxplot represents the 80\% prediction interval and the ends of the whiskers mark the 95\% prediction interval.  The corresponding observed average TFR based on the 2010 WPP is shown as a red dot. }
\label{avePlots}
\end{figure}

Figure \ref{avePlots} shows the posterior distribution of regional TFR for four regions, with the observed regional value shown in red.  The box associated with a given period and projection method represents the 80\% interval and the ends of the whiskers correspond to the 95\% interval.  
For Northern Europe the current model prediction intervals do not 
cover the observed value in 1995 and 2000, but the correlation model prediction intervals do.  Similar patterns are seen in the other regions: the prediction intervals based on the correlation model are wider, reflecting greater uncertainty, and contain more of the observed regional TFR values than the current model assuming independent errors.

Since the estimated correlations are larger when both countries have low TFR values, bigger differences between the current model and the correlation model prediction intervals are seen for regions like Northern Europe and Central Asia, for which the majority of the countries have completed most of the fertility decline.  Regions that have few countries with TFR less than 5, such as Eastern and Western Africa, showed little change between the prediction intervals from the current model and the correlation model, as expected. \\

\subsection{Effect of Taking Account of Correlation}
  
For a given parameter vector $\boldsymbol{\theta}$ and time period $t$, the effect of taking account of correlation on the variance of the regional TFR  
can be quantified analytically.  We denote by $p_{i}$ the proportion of the region's female population that resides in country $i$,
by $f_{i}$ the TFR in country $i$ in time period $t$, 
and by $N$ the number of countries in the region.
The regional weighted average TFR is then $\sum_{i} p_{i}f_{i}$ where the sum is over all countries in the region. 
 
If the forecast errors are assumed to be independent as in the current model, 
the predictive variance of the regional TFR in time period $t$ is 
$\text{Var}[\sum_{i} p_{i}f_{i}]  = \sum_{i} p_{i}^{2}  \text{Var}[f_{i}]$, 
where $\text{Var}[f_{i}]=\widetilde{\sigma}_{i,t}^{2}$.  As we project TFR into the future, eventually all countries will be in the last phase of the model, having completed their fertility transition, where $\text{Var}[f_{i}]=s^{2}$.  
When all countries in the region are in the post-transition phase,  
\begin{equation}
\text{Var} \left[\sum_{i} p_{i}f_{i} \right]  =  s^{2} \sum_{i} p_{i}^{2} \label{indvar}
\end{equation}
under the current model.  

We will refer to $\sum_{i} p_{i}^{2}$ as the independence factor (IF) 
since it represents the ratio of the regional variance to the 
country-specific variance assuming independence in the post-transition phase.
It indicates the effect of the distribution of the population across
countries in the region and shows that the more evenly the regional female population is spread amongst the countries within the region, the greater the variability in the regional estimate.  This comes from the fact that the IF 
is maximized when each $p_{i} = \frac{1}{N}$ for all countries in the region.

The variance of a region's TFR under the correlation model is 
$$\text{Var}\left[\sum_{i} p_{i}f_{i} \right]  =  \left( \sum_{i} p_{i}^{2} \text{Var}[f_{i}] + 2 \sum_{i < j} p_{i} p_{j} \sqrt{\text{Var}[f_{i}]\text{Var}[f_{j}]}R_{t}[i,j] \right).$$  
When all countries have completed the fertility transition, this becomes
 \begin{equation}
 \text{Var} \left[\sum_{i} p_{i}f_{i} \right]  =s^{2} \left( \sum_{i} p_{i}^{2} + 2 \sum_{i < j} p_{i} p_{j} R_{t}[i,j] \right). \label{corvar}
\end{equation}
We will refer to $\left( \sum_{i} p_{i}^{2} + 2 \sum_{i < j} p_{i} p_{j} R_{t}[i,j] \right)$ as the dependence factor (DF) since it is the multiplicative factor in the variance under the correlation model.  
Equation \eqref{corvar} shows that the larger the country correlations, 
especially between those countries with a relatively high proportion of 
the regional female population, the larger the variance of the regional TFR.  

The ratios of the dependence factor to the independence factor for the 22 UN 
regions are shown in Table \ref{varChanges}.  The regions with the largest 
ratios are those whose predictive distributions are most impacted by 
between-country correlations.  
For example, Western Africa and Eastern Asia's ratios are both greater than 2.5, indicating that the variance of the regional TFR predictive distributions is more than 2.5 times greater for the model with the correlation structure than that from the current model.  Those regions with ratios close to 1, such as Northern America and Australia/New Zealand, have similar predictive distributions from the two models.  

\begin{table}[ht!]
\begin{center}
\caption{The effect of the correlation model on the variance of the regional weighted average TFRs.  The ratio of the dependence factor to the independence factor (DF/IF) indicates the multiplicative increase in the variance of the regional TFR when using the correlation model compared to the current model where forecast errors are assumed independent.  The ``max proportion'' column shows the largest proportion of the current region female population that is attributed to a single country and N is the number of countries in the region.}
\label{varChanges}
\begin{tabular}{lrcr}
  \hline
 Region & DF/IF & Max Proportion & N \\ 
  \hline
 Northern America & 1.10 & 0.90 & 2 \\ 
 Eastern Asia & 1.09 & 0.85 & 8  \\ 
 Eastern Africa & 3.03 & 0.22 & 15  \\ 
 Middle Africa & 1.98 & 0.48 & 6  \\ 
 Northern Africa & 1.92 & 0.39 & 7  \\ 
 Southern Africa & 1.14 & 0.87 & 5  \\ 
 Western Africa & 1.43 & 0.59 & 13 \\ 
 Caribbean & 1.94 & 0.27 & 16  \\ 
 Central America & 1.25 & 0.73 & 8 \\ 
 South-Eastern Asia & 1.76 & 0.40 & 10 \\ 
  Western Asia & 2.57 & 0.33 & 18  \\ 
  Eastern Europe & 1.88 & 0.49 & 10  \\ 
 Northern Europe & 1.34 & 0.63 & 11  \\ 
  Southern Europe & 1.65 & 0.39 & 12 \\ 
 Western Europe & 1.91 & 0.43 & 7  \\ 
 Australia/New Zealand & 1.08 & 0.83 & 2 \\ 
 Melanesia & 1.12 & 0.78 & 5  \\ 
 South America & 1.95 & 0.50 & 13 \\ 
 Micronesia & 1.22 & 0.62 & 2  \\ 
 Polynesia & 1.33 & 0.48 & 3  \\ 
 Central Asia & 2.11 & 0.45 & 5  \\ 
 Southern Asia & 1.34 & 0.73 & 8 \\ 
   \hline
\end{tabular}
\end{center}
\end{table}

Both the between-country correlations and the proportion of the regional 
female population within each country influence the effect of the correlations.
The number of countries in the region and the proportion of the regional female population that live in the largest country are also shown in Table \ref{varChanges}.  If a high proportion of the female population lives in a single country, correlations will not have a large effect on the prediction intervals for the region.  Examples of this include Northern America and Eastern Asia which have low DF/IF ratios and high proportions of the female population in a single country.   
Overall, the regions whose predictive intervals in the future will 
be most highly affected by between-country correlations include Middle, Eastern and Northern Africa, Western and Central Asia, Eastern and Western Europe, South America, and the Caribbean.  

\subsection{Comparison to Previous Results}
Others have investigated the correlation between country forecast errors and obtained similar results to those reflected by the correlation model here.  
\cite{KeilmanPham2004} modeled TFR in 18 countries in the European Economic Area (EEA) with an autoregressive conditional heteroscedastic model and calculated the average correlation between country TFR errors to be 0.33.  Although not all countries in the EEA are within the same UN region, in our model two countries with low TFR that are in the same region and not contiguous have a correlation of 0.2 and those not in the same region but are contiguous have a value of 0.37.  
The magnitudes of these correlations are similar to those found by
\cite{KeilmanPham2004}.

When \cite{Alho2008} further studied the correlation matrix obtained by \cite{KeilmanPham2004}, he found a stark contrast between the correlations between the Mediterranean countries (Portugal, Spain, Italy, and Greece) and all others.  His estimate of the average correlation between forecast
errors in Mediterranean and non-Mediterranean countries 
was 0.12 and the correlation within each of these groups was 0.3.
Recall that in our model, the correlation between countries that have low TFR and have no geographic or colonial relation is 0.11.  
Overall the correlations between and within groups of countries from our model 
are comparable to those found empirically by others.

\cite{WilsonBell} modeled TFR using a random walk with drift and found the correlation between errors for Queensland and the rest of Australia to be 0.4.  
According to our correlation model, when TFR is less than 5, as it has been in Australia for many decades, the correlation between Australia and a hypothetical country contiguous to it would be 0.46.  
This is consistent with the result of \cite{WilsonBell}.  

\section{Discussion}
When producing probabilistic population projections for country aggregates, it is critical to take
account of between-country correlations in forecast errors of vital rates
\citep{Lutz1996,Lee1998,Beyond6}. In this paper we have proposed a method
for estimating between-country correlations in forecast errors of 
the TFR for all countries and using them to produce probabilistic TFR
forecasts for aggregates of countries such as regions.
For many country pairs there are few relevant data available,
and so we estimate the correlations by modeling them as a function
of three time-invariant predictors.

The resulting method yields the same probabilistic projections of TFR
for individual countries as the Bayesian hierarchical model of 
\cite{Leontine}, which was used by the UN for its (deterministic)
medium population projections for all countries in the 2010 WPP \citep{WPP}.
In an out-of-sample validation experiment, our correlation extension yielded better
coverage of predictive intervals than the current model of \cite{Leontine}
that does not explicitly take account of between-country correlations.
The posterior samples produced by our method can be incorporated into
probabilistic population projections in the same way as those produced
by the current method (Raftery et al 2012b). \nocite{Raftery&2012}

Other ways of doing this have been proposed. Based on \cite{AlhoSpencer}'s 
method of constructing correlated projections using random seeds, 
\cite{LutzSS1997} and \cite{LutzSS2001} combined projections, where the within and/or between region country correlation was zero or one, 
to obtain overall correlations of 0.5 and 0.7 between forecast errors.
Although this method produces forecast errors with the desired marginal 
correlation, the individual forecasts come from a mixture distribution 
of two extreme scenarios neither of which is realistic.

\cite{KeilmanPham2004} and \cite{Alho2008}
estimated correlations between TFR forecast errors for a set of
European countries for which long and high quality time series data
are available, and for which the TFRs have been low for a long time in 
most cases. This is the best case scenario, for which empirical
estimates of the correlations are reasonably accurate and
further modeling is probably unnecessary. Our method gives
similar estimates to theirs for the countries that they consider.
\cite{WilsonBell} developed probabilistic population projections for 
Queensland and the rest of Australia using an empirical correlation 
between TFR errors. Again, this is a good data situation, and
their empirical correlation estimates are consistent with our model-based ones.

\bibliographystyle{chicago}
\bibliography{20121130_FertAgg.bib}

\end{document}